%% file: main.tex
\documentclass{article} 
\usepackage{iclr2024_conference,times}

\input{math_commands.tex}

\usepackage{hyperref}
\usepackage{url}
\usepackage{amsmath}
\usepackage[english]{babel}
\usepackage{blindtext}
\usepackage{dsfont}
\usepackage{bbm}
\usepackage{amssymb}
\usepackage{algorithm}
\usepackage[noend]{algpseudocode}
\usepackage{booktabs} 
\usepackage{wrapfig}
\usepackage{graphicx}
\usepackage{caption, subcaption}
\usepackage{enumitem}
\usepackage{balance}

\input{defs}

\title{{Is the Watermarking of LLM-Generated \\ Code Robust?}}

\author{Tarun Suresh$^{1}$,\ \ Shubham Ugare$^{1}$,\ \ \textbf{Gagandeep Singh}$^{1, 2}$,\ \ Sasa Misailovic$^{1}$  \\
$^1$University of Illinois Urbana-Champaign, \ \ $^2$VMware Research\\
\{\texttt{tsuresh3}, \texttt{sugare2}, \texttt{ggnds}, \texttt{misailo}\}\texttt{@illinois.edu} \\
}

\iclrfinalcopy 
\begin{document}

\maketitle

\begin{abstract}
We present the first in-depth study on the robustness of existing watermarking techniques applied to code generated by large language models (LLMs). As LLMs increasingly contribute to software development, watermarking has emerged as a potential solution for detecting AI-generated code and mitigating misuse such as plagiarism or the automated generation of malicious programs. While previous research has demonstrated the resilience of watermarking in the text setting, our work reveals that watermarking techniques are significantly more fragile in code-based contexts. Specifically, we show that simple semantic-preserving transformations—such as variable renaming and dead code insertion—can effectively erase watermarks without altering the program’s functionality. To systematically evaluate watermark robustness, we develop an algorithm that traverses the Abstract Syntax Tree (AST) of a watermarked program and applies a sequence of randomized, semantics-preserving transformations. Our experimental results, conducted on Python code generated by different LLMs, indicate that even minor modifications can drastically reduce watermark detectability, with true-positive rates (TPR) dropping below 50\% in many cases. 

\vspace{0.1in}

Our code is publicly available at {\color{blue}\url{https://github.com/uiuc-arc/llm-code-watermark}}.


\end{abstract}

\section{Introduction}
\vspace{-.15in}
\label{intro}
\input{intro}

\section{Robustness of Watermarked Code}
\vspace{-0.1in}
\label{sec:method}

\subsection{LLM Watermarking}
\label{sec:background}
\input{background}

\subsection{Watermarked Program Transformation}
\label{sec:method_algo}

\input{method}

\vspace{-.1in}
\section{Experimental Methodology}
\vspace{-.1in}
\label{sec:experiment}
\input{experiment}

\vspace{-.05in}
\section{Experimental Results}
\vspace{-.1in}
\label{sec:evaluation}
\input{evaluation}

\vspace{-.05in}
\section{Related Work}
\vspace{-.1in}
\label{sec:related}

\input{related}

\section{Conclusion}
\label{sec:discussion}
\input{discussion}

\clearpage

\section*{ACKNOWLEDGMENTS}
We thank the anonymous reviewers for their comments. This research was supported in part by NSF Grants No. CCF-1846354, CCF-2217144, CCF-2238079, CCF-2313028, CCF-2316233, CNS-2148583, and Google Research Scholar award.

\bibliographystyle{iclr2024_conference}
\bibliography{ref}


\end{document}

%% file: math_commands.tex
\usepackage{amsmath,amsfonts,bm}

\def\eqref#1{equation~\ref{#1}}

\def\1{\bm{1}}

\DeclareMathAlphabet{\mathsfit}{\encodingdefault}{\sfdefault}{m}{sl}
\SetMathAlphabet{\mathsfit}{bold}{\encodingdefault}{\sfdefault}{bx}{n}



%% file: defs.tex
\newcommand{\tablesize}{\footnotesize}

\newcommand{\basea}{UMD}
\newcommand{\baseb}{Unigram}

%% file: intro.tex
The rapid advancement of large language models (LLMs) like GPT and Codex in understanding and generating code holds transformative potential for software development \citep{chen2021codex, openai2023gpt4, cornstack, swerank, assemblyrl, codearc, satbench, vericoder}, especially when combined with constrained decoding algorithms \citep{ugare2024syncode, itergen, crane, dingo}, which improve the syntactic and semantic correctness of the LLM output. 
However, the use of LLMs for coding tasks raises concerns about misuse such as code plagiarism and malware generation. 
Combating misuse requires accurate detection of LLM-generated code, which is challenging as LLMs are designed to produce realistic output that mimics human-generated code.


To address this issue, researchers have developed various \textit{watermarking techniques}, which inject hidden patterns in the generated output based on a hash or cryptographic key \citep{pmlr-v202-kirchenbauer23a, zhao2023provable, kuditipudi2023robust}. A critical challenge lies in potential human or automated modifications that can erase the patterns, undermining the watermark's detectability.


\noindent{\bf Motivation.} 
%
%
%
Previous studies have shown that at least 50\% of LLM-generated tokens need to be modified to remove a watermark \citep{kuditipudi2023robust}. In plain text, this task is inherently challenging, requiring extensive human paraphrasing or the use of another language model \citep{kirchenbauer23b}. On the other hand, code is significantly easier to modify. For instance, changes to one part of a program (e.g., renaming a variable), can impact the whole program. Likewise, semantic-preserving modifications like adding dead code or employing obfuscation do not alter program behavior, enabling adversaries to easily make significant changes without compromising code quality and thereby reducing the detectability of watermarks. 


\noindent{\bf This Work.} We are the first to investigate the robustness of watermarking Python code generated by LLMs. 
We propose an algorithm that walks the Abstract Syntax Tree (AST) of the watermarked code and randomly applies semantic-preserving program modifications. We observe a significantly lower true-positive rate (TPR) of detection even under simple modifications, underscoring the need for robust LLM watermarks tailored specifically for code. 





     


%% file: background.tex


Let $x$ denote the sequence of tokens of length $m$. For an auto-regressive language model $M$, the objective of watermarking is to generate a watermarked completion $y^w$ given $x$ by embedding a hidden pattern based on a hash or cryptographic key $\zeta$. A detector can then check if $y^w$ is watermarked or not using $\zeta$. A watermarking scheme consists of the following two algorithms: 


$\bullet$ \textbf{Watermark\,($M$, $x$, $\zeta$)}: Let $p_{t} := \mathds{P}_{M(x)}\left[y_{t} = \cdot \mid y_{1: t- 1}\right]$ represent the conditional probability distribution over $V$ of the t-th token generated by $M$. 
The algorithm uses a function $\Gamma(\zeta, p_{t})$ that maps $\zeta$ and $p_{t}$ to a modified distribution $\hat{p}_{t}$ over the next token. 
Output $y^w$ generated by iteratively sampling $y^w_{t}$ from $\hat{p}_{t} = \Gamma(\zeta, p_{t})$ using any of the standard decoding techniques.
    
$\bullet$ \textbf{Detect\,($y$, $\zeta$)}: Given a completion $y$ and key $\zeta$, compute a p-value $p$ with respect to the null hypothesis that $y$ was generated independently of $\zeta$. Return $\mathbbm{1}_{p \leq p_\textit{threshold}}$.

%% file: method.tex


\input{algos.tex/algo1}

In practice, a user may transform a watermarked code $y^w\sim\text{Watermark($M$, $x$, $\zeta$)}$ into a semantically equivalent $y_{A}$ such that $\text{Detect}(y_{A}, \zeta) = 0$. We consider that the user has only black-box input-output access to $M$ and has no knowledge of the watermarking algorithm, $\zeta$, or the detection threshold. 
The user can apply a series of $d$ semantic-preserving transformations $\{ T_{1}, T_{2}, \dots, T_{d}\}$, e.g., inserting print statements or renaming {variables, to modify the code.}


We replicate these realistic program modifications in Algorithm~\ref{alg:simple_attack}. The algorithm takes the watermarked code $y^w$ and the number of transformations $d$ to apply as input. The algorithm parses $y^w$ to obtain the $AST$ representation of the code. At each iteration $k$, a transform $T_k$ is selected at random. The algorithm traverses the $AST$ to determine the set of all possible insertion, deletion, or substitution locations $\mathcal{S}$ for $T_{k}$. It then transforms $AST$ at a randomly selected $s \sim \mathcal{S}$ subtree by $T_{k}$ by replacing the subsequence of terminals with a "hole" and then completing it with a random syntactically-valid sequence $\eta_{k} \sim \Sigma^*$.

%% file: algos.tex/algo1.tex
\begin{wrapfigure}{R}{0.45\textwidth}
\vspace{-.3in}
\begin{minipage}{0.45\textwidth}
\begin{algorithm}[H]
\footnotesize
\caption{Watermarked Program Transformation}
\label{alg:simple_attack}
\textbf{Inputs:} $y^w$: watermarked code, $\mathcal{T}$: set of transformations, $d$: number of transformations to apply
\begin{algorithmic}[1]
\Function{Perturb}{$y^w$, $d$, $\mathcal{T}$}
\State $AST \gets \text{parse}(y^w)$
\For{$k \gets 1$ to $d$}
\State $T_{k} \sim \mathcal{T}$
\State $\mathcal{S} \gets \text{visit}(AST, T_{k})$
\State $s \sim \mathcal{S}$; $\eta_{k} \sim \Sigma^*$
\State $AST \gets \text{transform}(AST, T_{k}, s, \eta_{k})$

\EndFor
\State \Return $\text{convertToCode}(AST)$
\EndFunction
\end{algorithmic}
\end{algorithm}
\end{minipage}
\vspace{-.3in}
\end{wrapfigure}

%% file: experiment.tex



\subsection{Experimental Setup}
\label{app:exp_setup}
We ran experiments on a 48-core Intel Xeon Silver 4214R CPU with 2 NVidia RTX A5000 GPUs. 
Our program transformations are implemented using the Python LibCST library. In our experiments, we sequentially apply each program transformation d = 5 times on the watermarked code.

\input{tables.tex/tab3}
Let $\Sigma$ denote the vocabulary of words that can be inserted or substituted into the program. In our implementation, we use the NLTK WordNet Python library for the vocabulary \citep{miller-1994-wordnet}.

\subsection{Semantic-Preserving Transformations}
We implement the following semantic-preserving transformations:
\newline
\textbf{AddDeadCode} A dead-code statement of the form {$i_1$ = rand() if ($i_1$ != $i_1$): $i_2$ = 0} is inserted at some random location in the program where $i_1, i_2 \sim \Sigma^*$. We sample $i_1, i_2$ until we get a valid variable name. 

\textbf{Rename} A single, randomly selected, formal identifier in the target program has its name replaced by a random word $i \sim \Sigma^*$. We sample $i$ until we get a valid variable name.

\textbf{InsertPrint} A single print statement {print($i$)}, is inserted at some random location in the program where $i$ is a string sequence of words such that $i \sim \Sigma^*$.

\textbf{WrapTryCatch} A random statement in the program is wrapped by a try-catch block. 

\textbf{Mixed} Apply $d$ of the aforementioned randomly selected transformations with replacement.

Table \ref{tab:prop_tokens} shows the mean proportion of tokens changed after applying each transformation. Each AddDeadCode and WrapTryCatch transformation modifies over 20\% of the tokens in the code. Consequently, in practice, it is extremely easy for the watermark to be erased away after even a few modifications by the user.

\subsection{Watermark Baselines}
\label{app:baselines}
Denote $|x|_G$ as the number of green list tokens for a generated text with length $T$. We experiment with the following two popular watermark schemes.
\begin{itemize}
    \item UMD~\citep{pmlr-v202-kirchenbauer23a} involves selecting a randomized set of ``green'' tokens before a word is generated, and then biasing green tokens during sampling. Detection is performed using the \textit{one proportion z-test}, where $z=2\left(|x|_G-T / 2\right) / \sqrt{T}$. to evaluate the null hypothesis $H_0$:\textit{The text sequence is generated with no knowledge of the red list rule}.
    \item Unigram \citep{zhao2023provable} is proposed as a watermark robust to edit property. Detection is performed by calculating the z-statistic $z=\left(|x|_G-\gamma T\right) / \sqrt{T \gamma(1-\gamma)}$.
\end{itemize}

For both UMD and Unigram, we set $\gamma$ = 0.25, where $\gamma$ represents the fraction of the vocabulary included in the green list. For the Unigram watermark, we set the strength parameter $\delta=2$, where the larger $\delta$ is, the lower the quality of the watermarked LM output, but the easier it is to detect. We observe empirically each function completion only has around 100 tokens on average and the TPR $< 0.3$. To increase the number of tokens and thus the TPR, after generating the watermarked code completions, we select 3 random function completions at a time and run detection collectively on the 3 functions. We reject the null hypothesis if $z > 3$. We show that the TPR $> 0.70$ and the FPR $\leq 0.01$ for the baseline by adopting this approach.

%% file: tables.tex/tab3.tex
\begin{wraptable}{r}{.4\textwidth} 
    \tablesize
    \centering
    \caption{Watermark detectability results}
\begin{tabular}{lc}
\hline
 Transformation   &   Proportion of \\
                  &    tokens changed   \\
\hline
 AddDeadCode      &                       0.36  \\
 InsertPrint      &                       0.12 \\
 Rename           &                       0.05 \\
 WrapTryCatch     &                       0.27 \\
 Mixed     &                       0.22 \\
\hline
\end{tabular}
\label{tab:prop_tokens}
\end{wraptable}

%% file: evaluation.tex
\subsection{Detectability of Watermarked Code For LlamA-7B}

\input{tables.tex/tab1}
We generate Python code completions using the LlamA-7B \citep{touvron2023llama} on the HumanEval \citep{chen2021codex} dataset. We watermark the code with state-of-the-art algorithms UMD and Unigram \citep{zhao2023provable,pmlr-v202-kirchenbauer23a}. We sequentially apply each program transformation $d = 5$ times on the watermarked code. As our transformation procedure is randomized, we run this experiment 3 times and compute the average of the results.
Table~\ref{tab:main_results} presents our main results for LlamA-7B. 
It shows that the program transformations greatly corrupt watermark detectability. 
Even the simplest transformations InsertPrint and Rename reduce the TPR by at least 1.3x. Complex alterations (e.g., WrapTryCatch and AddDeadCode) reduce the TPR much more significantly. 

\subsection{Detectability of Watermarked Code For CodeLlamA-7B}
\label{app:eval_cl}
\input{tables.tex/tab2}
Similarly, we evaluate the detectability of watermarked Python code completions generated by CodeLlamA-7B \citep{touvron2023llama} on HumanEval and present the results in Table~\ref{tab:code_llama_results}. Similar to LlamA-7B, we observe that the program transformations greatly reduce watermark detectability. Simple transformations like InsertPrint and Rename reduce the TPR by at least 1.4x. We observe even larger reductions for more complex modifications (e.g., WrapTryCatch and AddDeadCode).

\subsection{ Varying the Number of Transformations}
\begin{figure*}[t]
    \vspace{-0.21in}
    \centering
    \begin{minipage}{0.48\linewidth}
        \centering\captionsetup[subfigure]{justification=centering}
        \includegraphics[width=\linewidth]{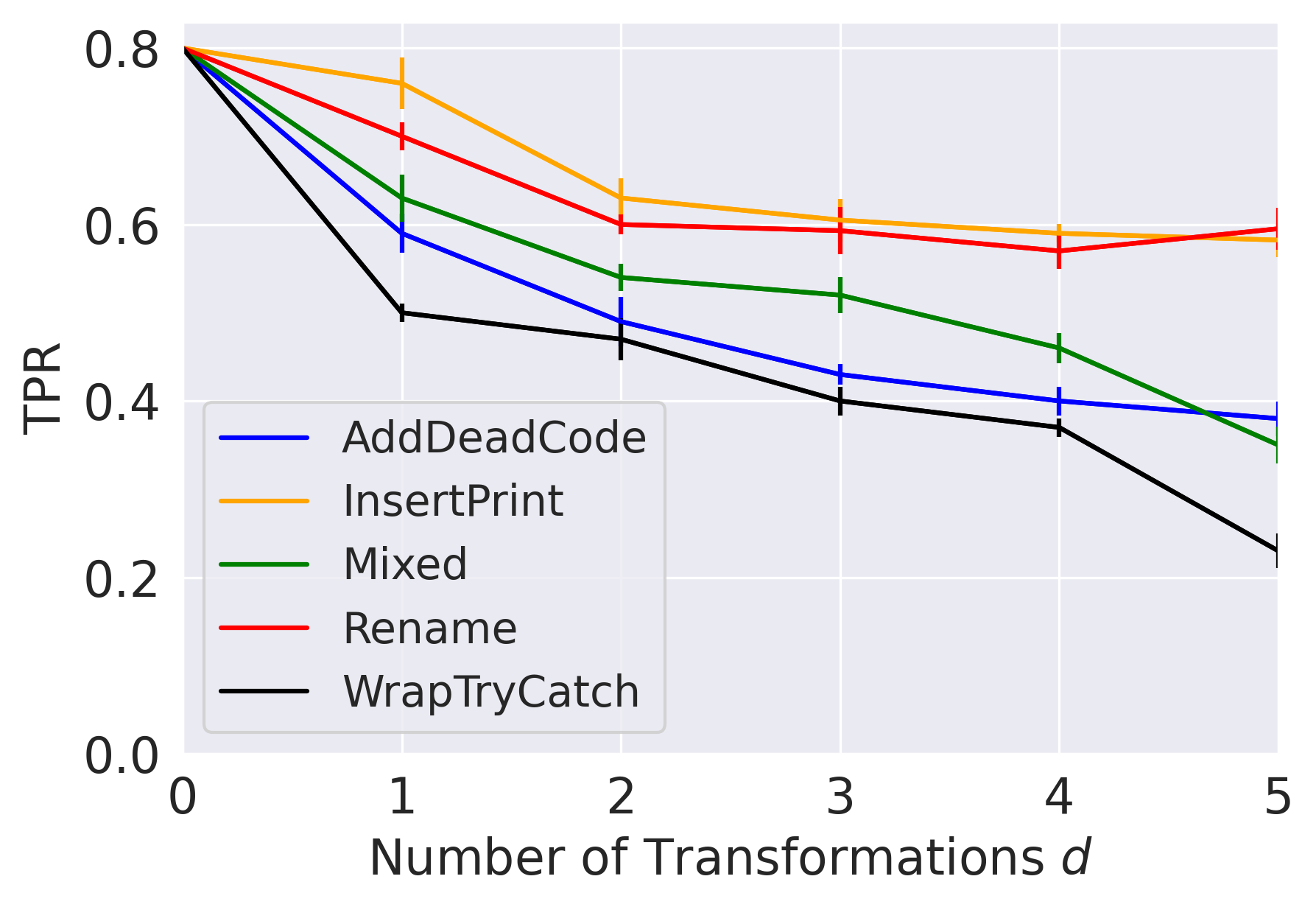}
        \subcaption{UMD}\label{subfig:umd}
    \end{minipage}%
    \hfill%
    \begin{minipage}{0.48\linewidth}
        \centering\captionsetup[subfigure]{justification=centering}
        \includegraphics[width=\linewidth]{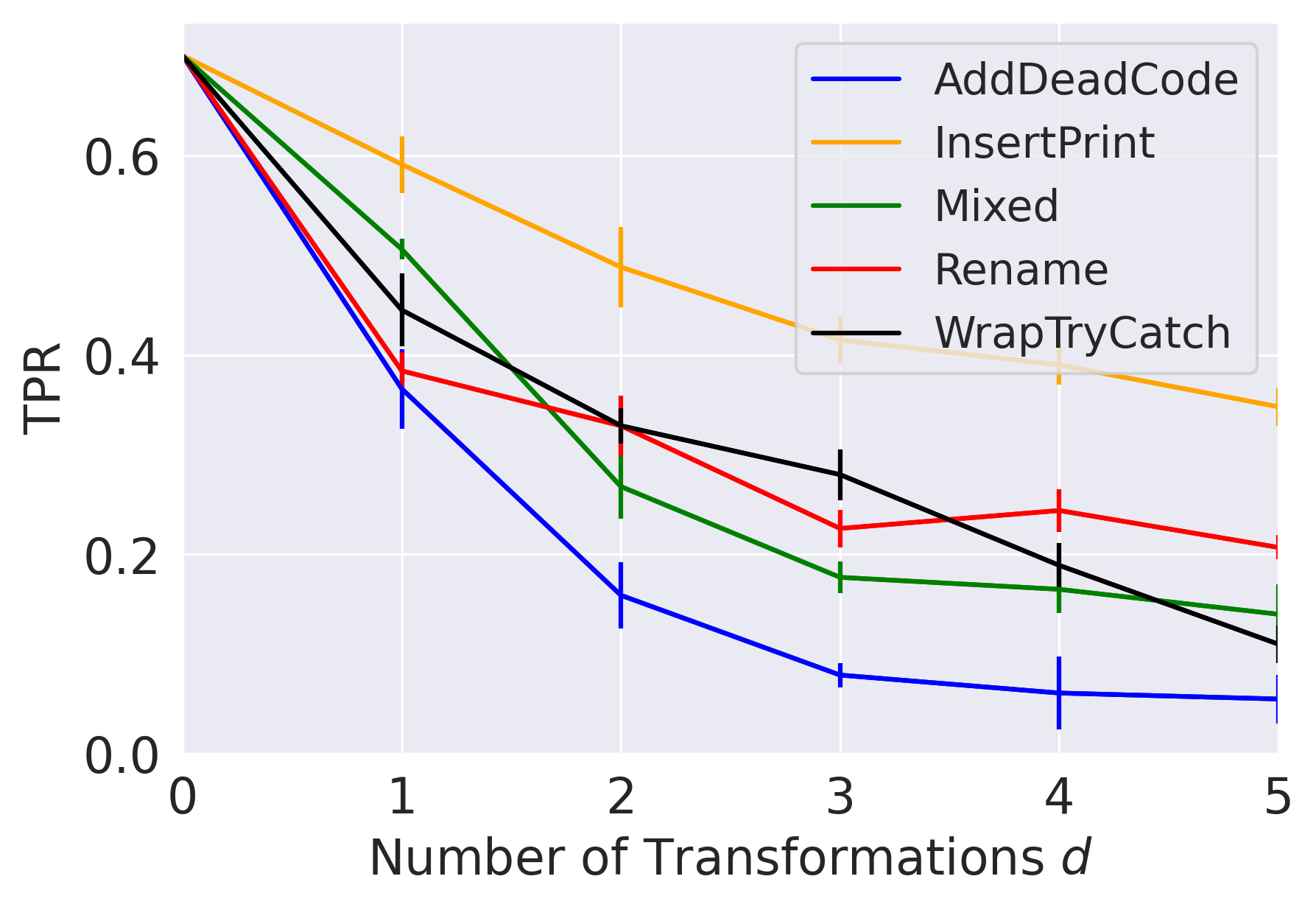}
        \subcaption{Unigram}\label{subfig:unigram}
    \end{minipage}
    \vspace{-0.07in}
    \caption{TPR vs the number of transformations $d$.}\label{fig:depth}
\end{figure*}

We further show the robustness of the watermark techniques by varying the number of modifications $d$ applied to the watermarked code from 0 to 5. 
Figure~\ref{fig:depth} shows that the TPR declines as $d$ increases. 
For instance, when employing 5 WrapTryCatch modifications, the TPR dropped to 0.22 for the UMD watermark and fell to 0.11 for the Unigram watermark. 
AddDeadCode and WrapTryCatch modifications exhibit a more pronounced impact on TPR, requiring fewer modifications to reduce TPR by over 2x compared to the other two modifications.



%% file: tables.tex/tab1.tex
\begin{wraptable}{r}{.46\textwidth} 
    \tablesize
    \centering
    \vspace{-0.15in}
    \caption{Llama-7B Watermark detectability results}
    \vspace{-0.1in}
    \begin{tabular}{@{}ll cc@{}}
        \toprule
        Algorithm & Transformation & \multicolumn{2}{c}{Detection Metrics} \\
        & & TPR & FPR \\
        \hline
        &  Original & 0.79 & 0 \\
         &  Rename & 0.57 & 0.01 \\
         \basea &  AddDeadCode &           0.38  & 0.06  \\
        &  InsertPrint & 0.58  & 0.06 \\
        & WrapTryCatch & 0.22 & 0.01\\
        & Mixed & 0.34 & 0.01\\
        \hline
        &  Original & 0.76 & 0.01 \\
         &  Rename & 0.20 &  0  \\
        \baseb &  AddDeadCode &  0.01 & 0 \\
        & InsertPrint & 0.32 & 0\\
        & WrapTryCatch & 0.11 & 0\\
        & Mixed & 0.14 & 0\\

        \bottomrule
    \end{tabular}
    \label{tab:main_results}
    \vspace{-.15in}
\end{wraptable}


%% file: tables.tex/tab2.tex
\begin{wraptable}{r}{.4\textwidth} 
    \tablesize
    \centering
    \vspace{-0.15in}
    \caption{CodeLlama-7B Watermark detectability results}
    \begin{tabular}{@{}ll rc@{}}
        \toprule
        Algorithm & Transformation & \multicolumn{2}{c}{Detection Metrics} \\
        & & TPR & FPR \\
        \hline
        &  Original & 0.82 & 0.01 \\
         &  Rename & 0.70 & 0.03\\
         \basea &  AddDeadCode &           0.45  & 0.04 \\
        &  InsertPrint & 0.61  & 0.04 \\
        & WrapTryCatch & 0.28 & 0.01\\
        & Mixed & 0.35 & 0.03\\
        \hline
        &  Original & 0.72 & 0.01\\
         &  Rename & 0.25 &  0 \\
        \baseb &  AddDeadCode &  0.03 & 0 \\
        & InsertPrint & 0.26 & 0 \\
        & WrapTryCatch & 0.09 & 0\\
        & Mixed & 0.12 & 0\\

        \bottomrule
    \end{tabular}
    \label{tab:code_llama_results}
    \vspace{-.15in}
\end{wraptable}


%% file: related.tex
\noindent{\bf LLM-generated text detection}
One approach to AI-generated text detection involves looking for features or statistical outliers that distinguish AI-generated text from human text. These features include entropy, perplexity, n-gram frequencies, rank, and, in the case of DetectGPT \citep{mitchell2023detectgpt}, the observation that minor perturbations of a LLM-generated text have lower log probability under the LLM on average than the original text. However, these zero-shot statistical detectors often require white-box access to model parameters, fail to detect texts generated by advanced LLMs, and rely on many text perturbations generated by another LLM, which is computationally expensive.

Another common approach is to train a binary classifier to distinguish between human and LLM-generated text. This approach assumes that LLM-generated text has distinguishing features that the trained model can identify. The fundamental problem with this is that generative models are designed with the intent of producing realistic output that is extremely hard to distinguish from that generated by humans. Specifically, recent advancements, including GPT-4 and other state-of-the-art models, are rapidly narrowing the gap between AI-generated and human-written text. As these generative models become more and more realistic, any black-box text distinguishers would incur large Type 1 and Type 2 errors. Distinguishers such as GPTZero \citep{tian2023gptzero}, Sniffer \citep{li2023origin}, and LMDNet \citep{wu2023llmdet} have no guarantee of correctness and are susceptible to issues such as out-of-distribution problems, adversarial attacks, and poisoning. 

\noindent{\bf LLM Watermarking Schemes}
Recently, \cite{pmlr-v202-kirchenbauer23a} gave the first LLM watermarking scheme with formal guarantees. Their watermark divides the vocabulary into a red list and a green list based on a hash of the previous tokens and biases sampling the next token from the green list during the decoding stage. Then, a detector can count the number of green list tokens and check whether this number is statistically significant to determine whether the model output is generated without knowledge of the red-green rule. 

In practice, the text generated by a language model is likely modified by a user before being fed to a detector. As a result, a line of work has focused on designing robust watermarks for text that are detectable even if the original LLM output was changed. For instance, \cite{zhao2023provable} simplify the soft watermarking scheme by consistently using a fixed red-green split and demonstrate that this new watermark is twice as robust to modifications as the baseline. Kirchenbauer et al discuss more robust detection schemes for when watermarked text is embedded in a larger human-written document. Additionally, \cite{kuditipudi2023robust} propose a watermarking scheme that uses a key that is as large as the LLM-generated text and then aligns that key with the text to compute an alignment cost. Recently, \cite{christ2023undetectable} and \cite{cryptoeprint:2023/1661} proposed cryptographic watermarking schemes for text that achieve robustness properties. 

However, these works focus mainly on watermarking LLM-generated text. They do not evaluate or provide formal guarantees on the performance of watermarks for LLM-generated code. Recently, \cite{lee2023wrote} proposed a new approach to watermark to LLM-generated code. They noticed that the performance of existing watermarking approaches does not transfer well to code generation tasks and attributed this to the fact that the entropy in the code generation is lower compared to that of plain text generation. They proposed a watermarking scheme called Selective Watermarking via Entropy Thresholding (SWEET) that only watermarks tokens with high enough entropy given a threshold. However, to compute the entropy of tokens at the time of detection, SWEET requires re-generating the entire code completion using the language model, which is computationally expensive. 

\noindent{\bf DNN Robustness}
A large body of research has focused on the robustness of LLMs and other DNNs against adversarial attacks (\citealp{rabbit, tarllm, zou2023universal, ugare2024incremental, zhang2023certified, 10.1145/3563324, Ugare_2023}). This line of work is orthogonal to our investigation as we instead focus on the robustness of whether LLM-generated output (code) can be reliably detected.

%% file: discussion.tex
We are the first to study the robustness of existing watermark techniques for LLM-generated Python code. We demonstrate that realistic program modifications can easily corrupt watermark detectability. We urge future work to develop resilient detection schemes for LLM-generated code, potentially by watermarking the syntax tree of the generated code, ensuring code quality, security, and reliability in the rapidly evolving landscape of LLMs.

%% file: main.bbl
\begin{thebibliography}{32}
\providecommand{\natexlab}[1]{#1}
\providecommand{\url}[1]{\texttt{#1}}
\expandafter\ifx\csname urlstyle\endcsname\relax
  \providecommand{\doi}[1]{doi: #1}\else
  \providecommand{\doi}{doi: \begingroup \urlstyle{rm}\Url}\fi

\bibitem[Banerjee et~al.(2025)Banerjee, Suresh, Ugare, Misailovic, and Singh]{crane}
Debangshu Banerjee, Tarun Suresh, Shubham Ugare, Sasa Misailovic, and Gagandeep Singh.
\newblock Crane: Reasoning with constrained llm generation, 2025.
\newblock URL \url{https://arxiv.org/abs/2502.09061}.

\bibitem[Chen et~al.(2021)Chen, Tworek, Jun, Yuan, de~Oliveira~Pinto, Kaplan, Edwards, Burda, Joseph, Brockman, Ray, Puri, Krueger, Petrov, Khlaaf, Sastry, Mishkin, Chan, Gray, Ryder, Pavlov, Power, Kaiser, Bavarian, Winter, Tillet, Such, Cummings, Plappert, Chantzis, Barnes, Herbert-Voss, Guss, Nichol, Paino, Tezak, Tang, Babuschkin, Balaji, Jain, Saunders, Hesse, Carr, Leike, Achiam, Misra, Morikawa, Radford, Knight, Brundage, Murati, Mayer, Welinder, McGrew, Amodei, McCandlish, Sutskever, and Zaremba]{chen2021codex}
Mark Chen, Jerry Tworek, Heewoo Jun, Qiming Yuan, Henrique~Ponde de~Oliveira~Pinto, Jared Kaplan, Harri Edwards, Yuri Burda, Nicholas Joseph, Greg Brockman, Alex Ray, Raul Puri, Gretchen Krueger, Michael Petrov, Heidy Khlaaf, Girish Sastry, Pamela Mishkin, Brooke Chan, Scott Gray, Nick Ryder, Mikhail Pavlov, Alethea Power, Lukasz Kaiser, Mohammad Bavarian, Clemens Winter, Philippe Tillet, Felipe~Petroski Such, Dave Cummings, Matthias Plappert, Fotios Chantzis, Elizabeth Barnes, Ariel Herbert-Voss, William~Hebgen Guss, Alex Nichol, Alex Paino, Nikolas Tezak, Jie Tang, Igor Babuschkin, Suchir Balaji, Shantanu Jain, William Saunders, Christopher Hesse, Andrew~N. Carr, Jan Leike, Josh Achiam, Vedant Misra, Evan Morikawa, Alec Radford, Matthew Knight, Miles Brundage, Mira Murati, Katie Mayer, Peter Welinder, Bob McGrew, Dario Amodei, Sam McCandlish, Ilya Sutskever, and Wojciech Zaremba.
\newblock Evaluating large language models trained on code.
\newblock 2021.

\bibitem[Christ et~al.(2023)Christ, Gunn, and Zamir]{christ2023undetectable}
Miranda Christ, Sam Gunn, and Or~Zamir.
\newblock Undetectable watermarks for language models, 2023.

\bibitem[Fairoze et~al.(2023)Fairoze, Garg, Jha, Mahloujifar, Mahmoody, and Wang]{cryptoeprint:2023/1661}
Jaiden Fairoze, Sanjam Garg, Somesh Jha, Saeed Mahloujifar, Mohammad Mahmoody, and Mingyuan Wang.
\newblock Publicly detectable watermarking for language models.
\newblock Cryptology ePrint Archive, Paper 2023/1661, 2023.
\newblock \_url{https://eprint.iacr.org/2023/1661}.

\bibitem[Kirchenbauer et~al.(2023{\natexlab{a}})Kirchenbauer, Geiping, Wen, Katz, Miers, and Goldstein]{pmlr-v202-kirchenbauer23a}
John Kirchenbauer, Jonas Geiping, Yuxin Wen, Jonathan Katz, Ian Miers, and Tom Goldstein.
\newblock A watermark for large language models.
\newblock In \emph{International Conference on Machine Learning}, 2023{\natexlab{a}}.

\bibitem[Kirchenbauer et~al.(2023{\natexlab{b}})Kirchenbauer, Geiping, Wen, Shu, Saifullah, Kong, Fernando, Saha, Goldblum, and Goldstein]{kirchenbauer23b}
John Kirchenbauer, Jonas Geiping, Yuxin Wen, Manli Shu, Khalid Saifullah, Kezhi Kong, Kasun Fernando, Aniruddha Saha, Micah Goldblum, and Tom Goldstein.
\newblock On the reliability of watermarks for large language models, 2023{\natexlab{b}}.

\bibitem[Kuditipudi et~al.(2023)Kuditipudi, Thickstun, Hashimoto, and Liang]{kuditipudi2023robust}
Rohith Kuditipudi, John Thickstun, Tatsunori Hashimoto, and Percy Liang.
\newblock Robust distortion-free watermarks for language models, 2023.

\bibitem[Laurel et~al.(2022)Laurel, Yang, Ugare, Nagel, Singh, and Misailovic]{10.1145/3563324}
Jacob Laurel, Rem Yang, Shubham Ugare, Robert Nagel, Gagandeep Singh, and Sasa Misailovic.
\newblock A general construction for abstract interpretation of higher-order automatic differentiation.
\newblock \emph{Proc. ACM Program. Lang.}, 6\penalty0 (OOPSLA2), oct 2022.
\newblock \doi{10.1145/3563324}.
\newblock URL \url{https://doi.org/10.1145/3563324}.

\bibitem[Lee et~al.(2023)Lee, Hong, Ahn, Hong, Lee, Yun, Shin, and Kim]{lee2023wrote}
Taehyun Lee, Seokhee Hong, Jaewoo Ahn, Ilgee Hong, Hwaran Lee, Sangdoo Yun, Jamin Shin, and Gunhee Kim.
\newblock Who wrote this code? watermarking for code generation, 2023.

\bibitem[Li et~al.(2023)Li, Wang, Ren, Sun, and Qiu]{li2023origin}
Linyang Li, Pengyu Wang, Ke~Ren, Tianxiang Sun, and Xipeng Qiu.
\newblock Origin tracing and detecting of llms, 2023.

\bibitem[Miller(1994)]{miller-1994-wordnet}
George~A. Miller.
\newblock {W}ord{N}et: A lexical database for {E}nglish.
\newblock In \emph{{H}uman {L}anguage {T}echnology: Proceedings of a Workshop held at {P}lainsboro, {N}ew {J}ersey, {M}arch 8-11, 1994}, 1994.

\bibitem[Mitchell et~al.(2023)Mitchell, Lee, Khazatsky, Manning, and Finn]{mitchell2023detectgpt}
Eric Mitchell, Yoonho Lee, Alexander Khazatsky, Christopher~D. Manning, and Chelsea Finn.
\newblock Detectgpt: Zero-shot machine-generated text detection using probability curvature, 2023.

\bibitem[OpenAI(2023)]{openai2023gpt4}
OpenAI.
\newblock Gpt-4 technical report, 2023.

\bibitem[Reddy et~al.(2025)Reddy, Suresh, Doo, Liu, Nguyen, Zhou, Yavuz, Xiong, Ji, and Joty]{swerank}
Revanth~Gangi Reddy, Tarun Suresh, JaeHyeok Doo, Ye~Liu, Xuan~Phi Nguyen, Yingbo Zhou, Semih Yavuz, Caiming Xiong, Heng Ji, and Shafiq Joty.
\newblock Swerank: Software issue localization with code ranking, 2025.
\newblock URL \url{https://arxiv.org/abs/2505.07849}.

\bibitem[Suresh et~al.(2024{\natexlab{a}})Suresh, Banerjee, and Singh]{rabbit}
Tarun Suresh, Debangshu Banerjee, and Gagandeep Singh.
\newblock Relational verification leaps forward with rabbit.
\newblock In A.~Globerson, L.~Mackey, D.~Belgrave, A.~Fan, U.~Paquet, J.~Tomczak, and C.~Zhang (eds.), \emph{Advances in Neural Information Processing Systems}, volume~37, pp.\  123328--123352. Curran Associates, Inc., 2024{\natexlab{a}}.
\newblock URL \url{https://proceedings.neurips.cc/paper_files/paper/2024/file/df15617fd6d6f78fcd485401d0598761-Paper-Conference.pdf}.

\bibitem[Suresh et~al.(2024{\natexlab{b}})Suresh, Reddy, Xu, Nussbaum, Mulyar, Duderstadt, and Ji]{cornstack}
Tarun Suresh, Revanth~Gangi Reddy, Yifei Xu, Zach Nussbaum, Andriy Mulyar, Brandon Duderstadt, and Heng Ji.
\newblock Cornstack: High-quality contrastive data for better code ranking, 2024{\natexlab{b}}.
\newblock URL \url{https://arxiv.org/abs/2412.01007}.

\bibitem[Suresh et~al.(2025)Suresh, Banerjee, Ugare, Misailovic, and Singh]{dingo}
Tarun Suresh, Debangshu Banerjee, Shubham Ugare, Sasa Misailovic, and Gagandeep Singh.
\newblock Dingo: Constrained inference for diffusion llms, 2025.
\newblock URL \url{https://arxiv.org/abs/2505.23061}.

\bibitem[Tamirisa et~al.(2025)Tamirisa, Bharathi, Phan, Zhou, Gatti, Suresh, Lin, Wang, Wang, Arel, Zou, Song, Li, Hendrycks, and Mazeika]{tarllm}
Rishub Tamirisa, Bhrugu Bharathi, Long Phan, Andy Zhou, Alice Gatti, Tarun Suresh, Maxwell Lin, Justin Wang, Rowan Wang, Ron Arel, Andy Zou, Dawn Song, Bo~Li, Dan Hendrycks, and Mantas Mazeika.
\newblock Tamper-resistant safeguards for open-weight llms, 2025.
\newblock URL \url{https://arxiv.org/abs/2408.00761}.

\bibitem[Tian \& Cui(2023)Tian and Cui]{tian2023gptzero}
Edward Tian and Alexander Cui.
\newblock Gptzero: Towards detection of ai-generated text using zero-shot and supervised methods, 2023.

\bibitem[Touvron et~al.(2023)Touvron, Lavril, Izacard, Martinet, Lachaux, Lacroix, Rozière, Goyal, Hambro, Azhar, Rodriguez, Joulin, Grave, and Lample]{touvron2023llama}
Hugo Touvron, Thibaut Lavril, Gautier Izacard, Xavier Martinet, Marie-Anne Lachaux, Timothée Lacroix, Baptiste Rozière, Naman Goyal, Eric Hambro, Faisal Azhar, Aurelien Rodriguez, Armand Joulin, Edouard Grave, and Guillaume Lample.
\newblock Llama: Open and efficient foundation language models, 2023.

\bibitem[Ugare et~al.(2023)Ugare, Banerjee, Misailovic, and Singh]{Ugare_2023}
Shubham Ugare, Debangshu Banerjee, Sasa Misailovic, and Gagandeep Singh.
\newblock Incremental verification of neural networks.
\newblock volume~7, pp.\  1920–1945. Association for Computing Machinery (ACM), June 2023.
\newblock \doi{10.1145/3591299}.
\newblock URL \url{http://dx.doi.org/10.1145/3591299}.

\bibitem[Ugare et~al.(2024{\natexlab{a}})Ugare, Gumaste, Suresh, Singh, and Misailovic]{itergen}
Shubham Ugare, Rohan Gumaste, Tarun Suresh, Gagandeep Singh, and Sasa Misailovic.
\newblock Itergen: Iterative structured llm generation, 2024{\natexlab{a}}.
\newblock URL \url{https://arxiv.org/abs/2410.07295}.

\bibitem[Ugare et~al.(2024{\natexlab{b}})Ugare, Suresh, Banerjee, Singh, and Misailovic]{ugare2024incremental}
Shubham Ugare, Tarun Suresh, Debangshu Banerjee, Gagandeep Singh, and Sasa Misailovic.
\newblock Incremental randomized smoothing certification.
\newblock 2024{\natexlab{b}}.

\bibitem[Ugare et~al.(2024{\natexlab{c}})Ugare, Suresh, Kang, Misailovic, and Singh]{ugare2024syncode}
Shubham Ugare, Tarun Suresh, Hangoo Kang, Sasa Misailovic, and Gagandeep Singh.
\newblock Syncode: Llm generation with grammar augmentation.
\newblock 2024{\natexlab{c}}.

\bibitem[Wei et~al.(2025{\natexlab{a}})Wei, Suresh, Cao, Kannan, Wu, Yan, Teixeira, Wang, and Aiken]{codearc}
Anjiang Wei, Tarun Suresh, Jiannan Cao, Naveen Kannan, Yuheng Wu, Kai Yan, Thiago S. F.~X. Teixeira, Ke~Wang, and Alex Aiken.
\newblock Codearc: Benchmarking reasoning capabilities of llm agents for inductive program synthesis, 2025{\natexlab{a}}.
\newblock URL \url{https://arxiv.org/abs/2503.23145}.

\bibitem[Wei et~al.(2025{\natexlab{b}})Wei, Suresh, Tan, Xu, Singh, Wang, and Aiken]{assemblyrl}
Anjiang Wei, Tarun Suresh, Huanmi Tan, Yinglun Xu, Gagandeep Singh, Ke~Wang, and Alex Aiken.
\newblock Improving assembly code performance with large language models via reinforcement learning, 2025{\natexlab{b}}.
\newblock URL \url{https://arxiv.org/abs/2505.11480}.

\bibitem[Wei et~al.(2025{\natexlab{c}})Wei, Tan, Suresh, Mendoza, Teixeira, Wang, Trippel, and Aiken]{vericoder}
Anjiang Wei, Huanmi Tan, Tarun Suresh, Daniel Mendoza, Thiago S. F.~X. Teixeira, Ke~Wang, Caroline Trippel, and Alex Aiken.
\newblock Vericoder: Enhancing llm-based rtl code generation through functional correctness validation, 2025{\natexlab{c}}.
\newblock URL \url{https://arxiv.org/abs/2504.15659}.

\bibitem[Wei et~al.(2025{\natexlab{d}})Wei, Wu, Wan, Suresh, Tan, Zhou, Koyejo, Wang, and Aiken]{satbench}
Anjiang Wei, Yuheng Wu, Yingjia Wan, Tarun Suresh, Huanmi Tan, Zhanke Zhou, Sanmi Koyejo, Ke~Wang, and Alex Aiken.
\newblock Satbench: Benchmarking llms' logical reasoning via automated puzzle generation from sat formulas, 2025{\natexlab{d}}.
\newblock URL \url{https://arxiv.org/abs/2505.14615}.

\bibitem[Wu et~al.(2023)Wu, Pang, Shen, Cheng, and Chua]{wu2023llmdet}
Kangxi Wu, Liang Pang, Huawei Shen, Xueqi Cheng, and Tat-Seng Chua.
\newblock Llmdet: A third party large language models generated text detection tool, 2023.

\bibitem[Zhang et~al.(2023)Zhang, Zhang, Hou, Fan, Li, Liu, Zhang, and Chang]{zhang2023certified}
Zhen Zhang, Guanhua Zhang, Bairu Hou, Wenqi Fan, Qing Li, Sijia Liu, Yang Zhang, and Shiyu Chang.
\newblock Certified robustness for large language models with self-denoising, 2023.

\bibitem[Zhao et~al.(2023)Zhao, Ananth, Li, and Wang]{zhao2023provable}
Xuandong Zhao, Prabhanjan Ananth, Lei Li, and Yu-Xiang Wang.
\newblock Provable robust watermarking for ai-generated text, 2023.

\bibitem[Zou et~al.(2023)Zou, Wang, Carlini, Nasr, Kolter, and Fredrikson]{zou2023universal}
Andy Zou, Zifan Wang, Nicholas Carlini, Milad Nasr, J.~Zico Kolter, and Matt Fredrikson.
\newblock Universal and transferable adversarial attacks on aligned language models, 2023.

\end{thebibliography}
